\documentclass[%
 reprint,
 nofootinbib,
 amsmath,amssymb,
 aps,
 pra,
 floatfix,
 twocolumn
]{revtex4-1}

\pdfoutput=1
\usepackage[spanish,english]{babel}
\usepackage[utf8]{inputenc}
\usepackage[T1]{fontenc}
\newcommand{\ignore}[1]{}
\usepackage{fancyhdr}
\usepackage{xcolor}
\usepackage{pgf}
\usepackage{graphicx}
\usepackage{appendix}
\usepackage{mathtools}
\usepackage{makecell}
\usepackage{multirow}
\usepackage[compact]{titlesec} 
\usepackage{physics}

\definecolor{mycolor}{RGB}{0, 0, 204}

\newcommand{\poughkeepsie}{\textit{poughkeepsie }}
\newcommand{\cnot}{\textsc{cnot} }
\newcommand{\eqn}[1]{Eq.~(\ref{#1})}

\begin{document}

\title{Modeling noisy quantum circuits using experimental characterization}
\author{Megan L. Dahlhauser}
\email{lillymn@ornl.gov}
\homepage{https://orcid.org/0000-0002-5747-9695}
\affiliation{Quantum Computing Institute, Oak Ridge National Laboratory, Oak Ridge, Tennessee, USA}
\affiliation{Bredesen Center for Interdisciplinary Research and Graduate Education, University of Tennessee, Knoxville, USA}
\author{Travis S. Humble}
\email{humblets@ornl.gov}
\homepage{https://orcid.org/0000-0002-9449-0498}
\thanks{\textit{Published by the American Physical Society under the terms of the Creative Commons Attribution 4.0 International license. Further distribution of this work must maintain attribution to the author(s) and the published article's title, journal citation, and DOI. Citation: Megan L. Dahlhauser and Travis S. Humble. Modeling noisy quantum circuits using experimental characterization. Phys. Rev. A, 103:042603, Apr 2021.} This manuscript has been authored by UT-Battelle, LLC under Contract No. DE-AC05-00OR22725 with the U.S. Department of Energy. The United States Government retains and the publisher, by accepting the article for publication, acknowledges that the United States Government retains a non-exclusive, paid-up, irrevocable, world-wide license to publish or reproduce the published form of this manuscript, or allow others to do so, for United States Government purposes. The Department of Energy will provide public access to these results of federally sponsored research in accordance with the DOE Public Access Plan. (http://energy.gov/downloads/doe-public-access-plan).}
\affiliation{Quantum Computing Institute, Oak Ridge National Laboratory, Oak Ridge, Tennessee, USA}
\affiliation{Bredesen Center for Interdisciplinary Research and Graduate Education, University of Tennessee, Knoxville, USA}

\begin{abstract}
Noisy intermediate-scale quantum (NISQ) devices offer unique platforms to test and evaluate the behavior of quantum computing. However, validating circuits on NISQ devices is difficult due to fluctuations in the underlying noise sources and other non-reproducible behaviors that generate computational errors. 
Here we present a test-driven approach 
that decomposes 
a noisy, application-specific circuit into a series of bootstrapped experiments on a NISQ device. By characterizing individual subcircuits, we generate a composite noise model for the original quantum circuit. 
We demonstrate this approach to model applications of GHZ-state preparation and the Bernstein-Vazirani algorithm on a family of superconducting transmon devices. We measure the model accuracy using the total variation distance between predicted and experimental results, and we demonstrate that the composite model works well across multiple circuit instances. Our approach is shown to be computationally efficient and offer a trade-off in model complexity that can be tailored to the desired predictive accuracy.
\keywords{Quantum Computing; Model and Simulation; Characterization}
\end{abstract}
\maketitle
\section{Introduction}
\label{sec:intro}
Quantum computing is a promising approach to accelerate computational workflows by solving problems with greater accuracy or using fewer resources as compared to conventional methods \cite{svore2016quantum,BrittISC2017,mccaskey2018language,riesebos2019quantum}. Testing and evaluation of early applications on experimental quantum processing units (QPUs) is now possible  using prototypes based on superconducting transmons \cite{Gambetta2017, Arute2019, Reagoreaao3603, PhysRevLett.122.110501} and trapped ions \cite{Figgatt2019, Monz1068, oxfordtrappedion, Kielpinski2002}  among other technologies. Although these QPUs lack the  fault-tolerant operations required for known computational speed ups, they offer the opportunity to understand the behaviors of noisy quantum computing  \cite{Preskill2018quantumcomputingin}. 
\par
Noisy, intermediate-scale quantum (NISQ) devices have enabled a wide range of early application demonstrations \cite{linke2017experimental,kandala2017hardware,dumitrescu2018cloud,klco2018quantum,mccaskey2019quantum,Arute2019}, but validating program performance in the presence of non-reproducible device behaviors remains a fundamental challenge. NISQ devices are characterized by noisy and erroneous operations, where gate characterizations often change in time and with the nature of the program being implemented \cite{VEITIA2019126131,rudinger2019probing}. The experimental characterization of individual gates has relied on high-fidelity physics models for the underlying devices with common methods including quantum state tomography (QST) \cite{Leibfried1996QST}, quantum process tomography (QPT)  \cite{poyatos1997complete,Chuang_1997QPT}, gate set tomography (GST) \cite{blumekohout2013robust}, and randomized benchmarking (RB) \cite{knill2008randomized, magesan2011scalable,PhysRevA.100.032304}. Physics-driven characterizations offer valuable insights into the underlying noise and errors that can inform the design of new devices and control pulses. However, translating from gate-level characterizations to circuit-level applications is typically resource intensive because these methods often scale exponentially with the size of the qubit register to be characterized. \cite{Eisert_2020}.
\par
As NISQ applications evolve toward deeper and wider quantum circuits, characterization methods must also extend to these larger scales. There is also a growing need for characterization techniques that can be executed swiftly and repeatedly to provide context-specific characterization data. Resource-intensive, physics-driven gate characterization techniques are not a scalable solution to characterizing devices and applications which are rapidly increasing in size and generally do not allow for a high level of dynamic tuning. Quantum circuit characterization methods may provide effective models of device behaviors that are efficient to generate and easy to interpret by a supporting programming environment, e.g., a compiler \cite{Dousti_2012, chong2017programming, shi2019optimized}. In particular, the validation of application behavior will require debugging methods and programming techniques that support mitigating computational errors in quantum circuits \cite{PhysRevA.100.032328,harper2019efficient}. Effective models of noisy gates and circuits have already informed robust programming methods that lead to increased application performance \cite{maciejewski2019mitigation,Tannu_2019,Murali:2019:NCM:3297858.3304075}, but a general method for composing noisy quantum circuit models is still needed.
\par
Here, we introduce methods for generating effective models for noisy quantum circuits in NISQ devices derived from experimental characterization. Our approach is based on modeling application-specific circuits using a suite of characterization tests that build a representative set of noisy subcircuit models. We compose noisy subcircuit models to generate noise models for more complicated circuits at larger scales, and we test the fidelity of the resulting model against experimental data. We show how to iteratively adjust the composite model selected for a noisy application circuit by comparing performance of the predicted behavior against application observations using the total variation distance (TVD) \cite{maciejewski2019mitigation}. The iterative and flexible nature of this modeling approach is demonstrated using applications based on GHZ-state preparation and the Bernstein-Vazirani algorithm for search. We develop model composition for the fixed-frequency superconducting transmon devices available from IBM, though we propose these techniques may extend to other NISQ devices as well. 
\par
This characterization method is a coarse-grained yet fast approach to characterization which scales linearly with the number of elements in the device, e.g. qubits and couplings. Furthermore, it allows for dynamic tuning of characterization data to every execution of a particular application and can be tailored to yield desired information, e.g. development of a noise model using depolarizing parameters or performance of an entangling gate creating an equal superposition. The tradeoff compared to physics-driven characterization techniques is less total information received, which in some cases may result in a lower accuracy in the final effective description of the device. 
\par
We present the steps in the modeling methodology in Sec.~\ref{sec:methods} followed by a series of  examples using the case of $n$-qubit GHZ states in Sec.~\ref{sec:application}. In Sec.~\ref{sec:qpus}, we present results from experimental characterization for the GHZ state on NISQ QPUs and discuss the role of model selection for characterization accuracy. In Sec.~\ref{sec:performance} we show the performance of our noise models composed from this characterization on the GHZ state experimental results. In Sec.~\ref{sec:bv}, we apply these models to the case of the $n$-bit Bernstein-Vazirani algorithm, while we offer final conclusion in Sec.~\ref{sec:fin}. 
\section{Model Selection Methodology}
\label{sec:methods}
\begin{figure}[ht]
\centering
\includegraphics[width=0.4\textwidth]{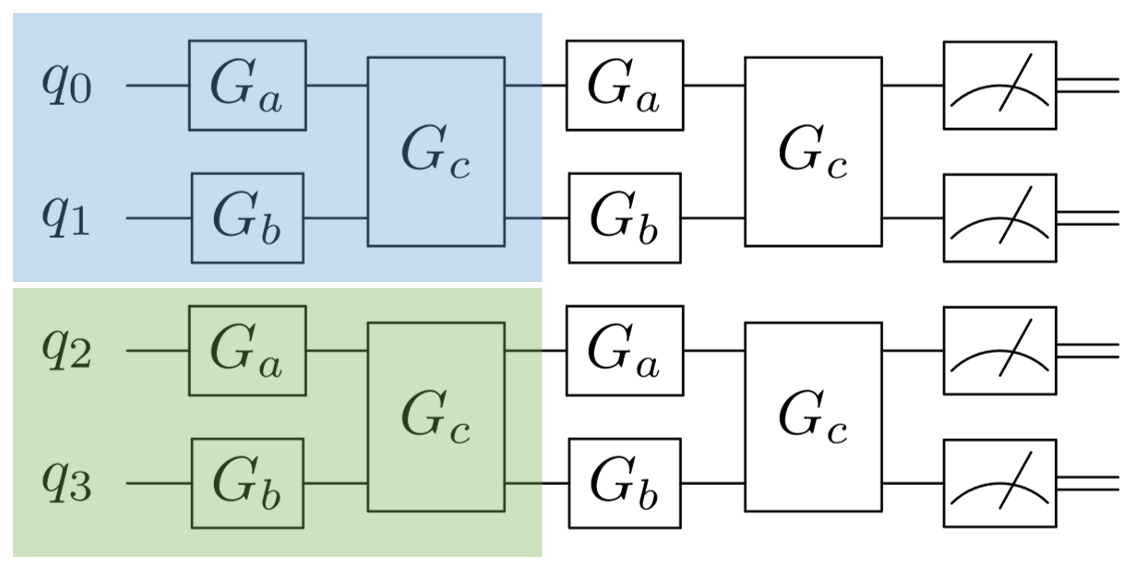}
\caption{An example of a subcircuit decomposition where subcircuit set $S=\{S_{blue},S_{green}\}$.}
\label{fig:example}
\end{figure}

We begin by detailing the coarse-grain modeling methodology before providing specific examples of its implementation. Consider the input for noisy circuit modeling to be an idealized quantum circuit $C$ that is expressed in the available instruction set architecture (ISA) for a given QPU \cite{BrittISC2017}. While the gates defined by the ISA may not be directly implemented within the QPU, the representation used for the ideal circuit will define the operators available for gate characterization. The input circuit is decomposed into a set $S(C) = \{S_{i} \}$ of idealized subcircuits $S_i$ that each represent a subsection of the total area of circuit $C$. The area of $C$ is defined by its width (register size) and depth (length of the operation sequence). The area of each subcircuit $S_i$ is defined by the selected subcircuit width taken from $C$ and the longest depth of the selected gate sequence. For example, a circuit $C$ composed of one- and two-qubit gates as shown in Fig.~\ref{fig:example} may be decomposed into a set $S$ of two-qubit subcircuits which have depth of two gates and width of two qubits.
\par
Circuit decomposition is not unique and a given decomposition is selected based on tradeoffs in the cost of characterizing each subcircuit, prior knowledge of the suspected device noise and error processes, and any potential structure or symmetry in the circuit design. A complete characterization requires every gate and register element within the input circuit to be included in at least one subcircuit. In general, the selected subcircuits need not be disjoint. The ability to tune the decomposition enables coarse-graining of the noisy circuit model, which is formed by composing the results from subcircuit characterization.
\par
Next we test each subcircuit to characterize the noise present within the coarse-grained area. Each test circuit specifies an idealized outcome based on the input state and gate sequence for the subcircuit instance. We select test circuits to be informative yet limited in both number and circuit dimensions in order to increase efficiency and improve scalability. To test a subcircuit $S_i$, we may select the full subcircuit $S_i$ provided the ideal outcome is known, but we may select additional test circuits to gain more information and refine our noise models. The set of test circuits $T = \{T_{i}\}$ is therefore at least as large as $S$ and generally larger. For example, given a two-qubit subcircuit $S_i$ consisting of a one-qubit gate followed by a two-qubit gate, we may select two test circuits--the first circuit consisting of the one-qubit gate and the second circuit consisting of both gates.
\par
The process for selecting test circuits $T(S)=\{T_i\}$ for each $S_i$ follows a set of guidelines detailed below.
\begin{enumerate}
    \item Identify the components used in $S_i$.
    \begin{enumerate}
        \item \label{decomp:qubits} Qubit register of size $n$ with qubit identities $q_j\in\{q_0, q_1, ..., q_n\}$
        \item \label{decomp:init} State preparation $\ket{\psi_j}$
        \item \label{decomp:meas} Measurement basis $B$
        \item \label{decomp:seq} Gate sequence $\mathcal{G}$
    \end{enumerate}
\item Generate measurement subcircuit $T_{meas}$ consisting of initialization of $\ket{\psi_j}$ and measurement in $B$ for each $q_j$. If $\ket{\psi_j}$ is unknown or more tests are needed, select or add the computational basis states $\ket{0}$ and $\ket{1}$. Additional input states may include superposition states such as $\ket{\psi}=(\ket{0}+\ket{1})/\sqrt{2}$ or randomly generated input states $\ket{\psi}=\alpha\ket{0}+\beta\ket{1}$.
\item Identify the set $g=\{g_k\}$ of the gates or gate compositions of $\mathcal{G}$ for which the expected outcomes may be calculated for a given input.
\item Select set $g'$ for testing. Elements of $g'$ are gates from $g$ or compositions of gates from $g$ which represent sequences of increasing depth from subcircuit $S_i$. The selection of $g'$ may be based on tradeoff in the cost of characterization or informed by prior knowledge of expected noise processes or iterative refinement, similar to subcircuit selection.
\item For each element $g'_k\in g'$, generate a circuit $T_k(g'_k)$ which consists of initialization of $\ket{\psi_j}$, application of $g'_k$ applied to the $q_j$ identified from $S_i$, and measurement in $B$.
\item The set of test circuits is $T=\{T_{meas}, T_i(g'_k) \hspace{0.1cm}\forall\hspace{0.1cm} g'_k\}$.
\end{enumerate}
\par
The implementation and execution of test circuits on a QPU generates a corresponding set of measurement observations. Each test circuit is executed multiple times to gather statistics from the distribution of results $R_{i}$ that characterize subcircuit $T_i$. The $i$-th characterization is denoted as $H_{i} = (T_{i}, R_{i})$ and the set of all characterizations is given as $H$. The number of characterizations is fixed by the number of test circuits $|T|$, while the number of measurement observations acquired for each test circuit is set by the sampling parameter $N_{s}$. Assuming the same sampling for all tests, then there are a total of $N_s |T|$ measurement observations, i.e., experiments, required for $H$.
\par
The results of experimental characterization are used to formulate concise approximate models of the subcircuits' observed behaviors. We model each noisy subcircuit as the idealized subcircuit followed by a quantum channel that accounts for the noise \cite{Bassi_2008}. Let the noisy subcircuit model $M_i = M(S_i, p_i)$ representing subcircuit $S_i$ depend on model parameters $p_i$. We estimate the channel parameters using the characterization $H_i$, where the method of parameter estimation will vary with the selected model. Parameter estimation may be either direct or optimized methods. For example, least-square error estimates may be used to estimate parameters from noisy measurement observations by optimizing the residual model error. 
\par
We quantify the error in the resulting models using the total variation distance (TVD) \cite{maciejewski2019mitigation}, which is defined as
\begin{equation}
    d_{\textsc{tv}}(H_i, M_i) = \frac{1}{2}\sum_k{\Big \lvert r^{(H_i)}(k) - r^{(M_i)}(k) \Big \rvert}
    \label{eq:TVD}
\end{equation}
where $r^{(H_i)}(k)$ is the probability of the $k$-th outcome of the test circuit $T_i$ and $r^{(M_i)}(k)$ is the corresponding probability predicted by the noisy circuit model.
The TVD vanishes as the predictions of the model become more accurate in reproducing the observed results and reaches a maximum of unity when the sets are completely disjoint.
\par
After estimating the model parameters $p = \{p_i\}$ for all subcircuits, the corresponding noisy circuit model $M(C, p)$ for the input circuit $C$ is composed. The method of composition of the noisy subcircuit models is paired with the decomposition method to ensure a consistent representation of the original input circuit. In the examples below, we consider modeling methods based on independent noisy subcircuit models that permit separable composition-decomposition methods and defer discussion of non-separable models, e.g., context-dependent noise, to Sec.~\ref{sec:fin}.
\par
Final selection of the noisy circuit model is then guided by the accuracy with which the composite model reproduces the performance of the circuit $C$ on the QPU. For clarity, we define the actual executed circuit $A = (C, R_{c})$ with $R_{c}$ the recorded results, and we measure the accuracy of the noisy circuit model as $d_{\textsc{tv}}(A, M)$. The desired TVD sets an upper bound on the threshold for model accuracy. If this user-defined threshold is not satisfied, selection of the noisy subcircuit models is revisited. This iteration may include refinement of the noisy subcircuit models to improve the accuracy of each $M_i$ or redefinition of the circuit composition-decomposition methods to manage the trade-offs in modeling complexity and accuracy. The former requires repeated post-processing analysis of the characterization $H$, whereas the latter requires additional characterization testing. In either case, model selection continues until the threshold has been meet. Once the accuracy threshold has been satisfied, noisy circuit modeling is complete.
\par
The noisy subcircuit models can then be tested for robustness in predicting the expected outcome from both the input circuit and other circuits executed on the characterized device. We again use TVD to measure the accuracy for selected models to characterize the behavior of other application circuits within the same QPU context.
\par
We summarize the complete procedure as follows.
\begin{enumerate}
    \item Identify ideal circuit $C$.
    \item \label{procedure:decomposition} Decompose the circuit into set $S(C)=\{S_i\}$ of ideal subcircuits $S_i$.
    \item \label{procedure:testcircuits} Select set of test circuits $T=\{T_i\}$ which define an input state and ideal outcome for each element in $S$.
    \item \label{procedure:noisemodel} Propose a noisy subcircuit model $M_i=M(S_i,p_i)$ for each element in $S$ parameterized by $p_i$.
    \item Implement and execute $T$ on QPU to generate experimental characterizations $H_i=(T_i,R_i)$ using results $R_i$ returned from QPU.
    \item Using set of characterizations $H=\{H_i\}$, fit noise parameters $p_i$ based on calculated expected probabilities for each $M_i$.
    \item \label{procedure:TVD} Compose the noisy circuit model $M(C,p)$ for the target circuit and compare the actual executed circuit $A=(C,R_C)$ with recorded results $R_C$ from the QPU to the noisy circuit model using $d_{TV}(A,M)$.
    \item \label{procedure:refine} If $d_{TV}$ is not at threshold return to \ref{procedure:decomposition}, apply refinements to \ref{procedure:decomposition}, \ref{procedure:testcircuits}, and \ref{procedure:noisemodel}, and continue to \ref{procedure:TVD} until threshold is met.
\end{enumerate}
\par
For step \ref{procedure:refine}, refinements to step \ref{procedure:decomposition} include additional elements selected from the set $g$, addition of compositions of elements in $g$ such that the test components are larger, or addition of elements to $g$ not explicitly represented in $\mathcal{G}$. Refinements to step \ref{procedure:testcircuits} include additional initializations as test circuits. Refinements to step \ref{procedure:noisemodel} include additional noise model parameters $p_i$ or different noise channels to define $M$.
\section{Application to GHZ States}
\label{sec:application}
\begin{figure}[t]
\centering
\includegraphics[width=0.4\textwidth]{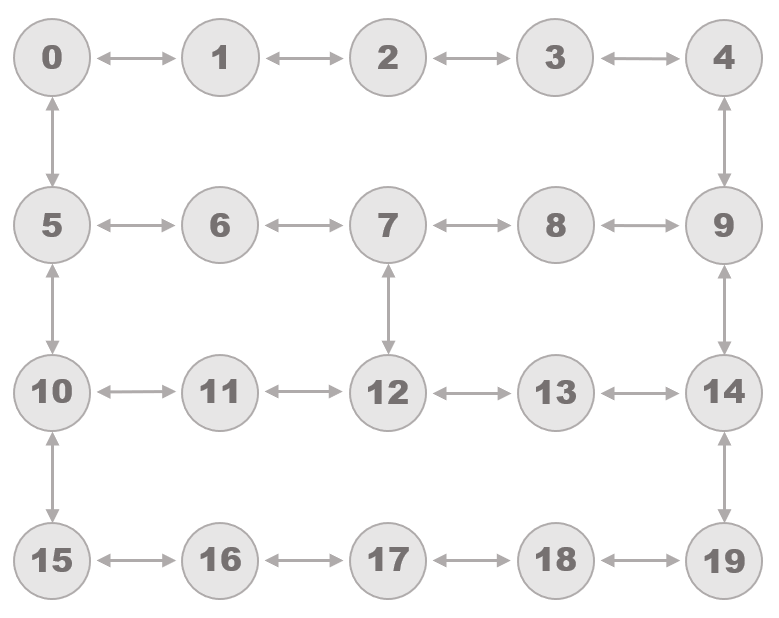}
\caption{A graphical representation of the register connectivity in the  \poughkeepsie QPU at the time of data collection, in which each node corresponds to a register element and directional edges indicate the availability of a programmable two-qubit cross-resonance gate.}
\label{fig:layout}
\end{figure}

We next illustrate the methodology of Sec.~\ref{sec:methods} using the example of a GHZ-state preparation and measurement circuit. We generate noisy quantum circuit models for this application for various circuit sizes executed on the IBM \poughkeepsie QPU, which has a register and layout as shown in Fig.~\ref{fig:layout}. All data for characterization tests and applications is collected in a single job sent to \textit{poughkeepsie}, a process which typically required under 30 minutes of execution time after queuing. As the \poughkeepsie device is periodically calibrated, our experimental demonstrations ensure that all data is collected within one calibration window to preserve the QPU context. The software implementation of our examples below as well as all experiment and simulation details such as subcircuits and noise models is available publicly \cite{publicrepo}.
\par
\begin{figure}[ht]
\centering
\includegraphics[width=0.45\textwidth]{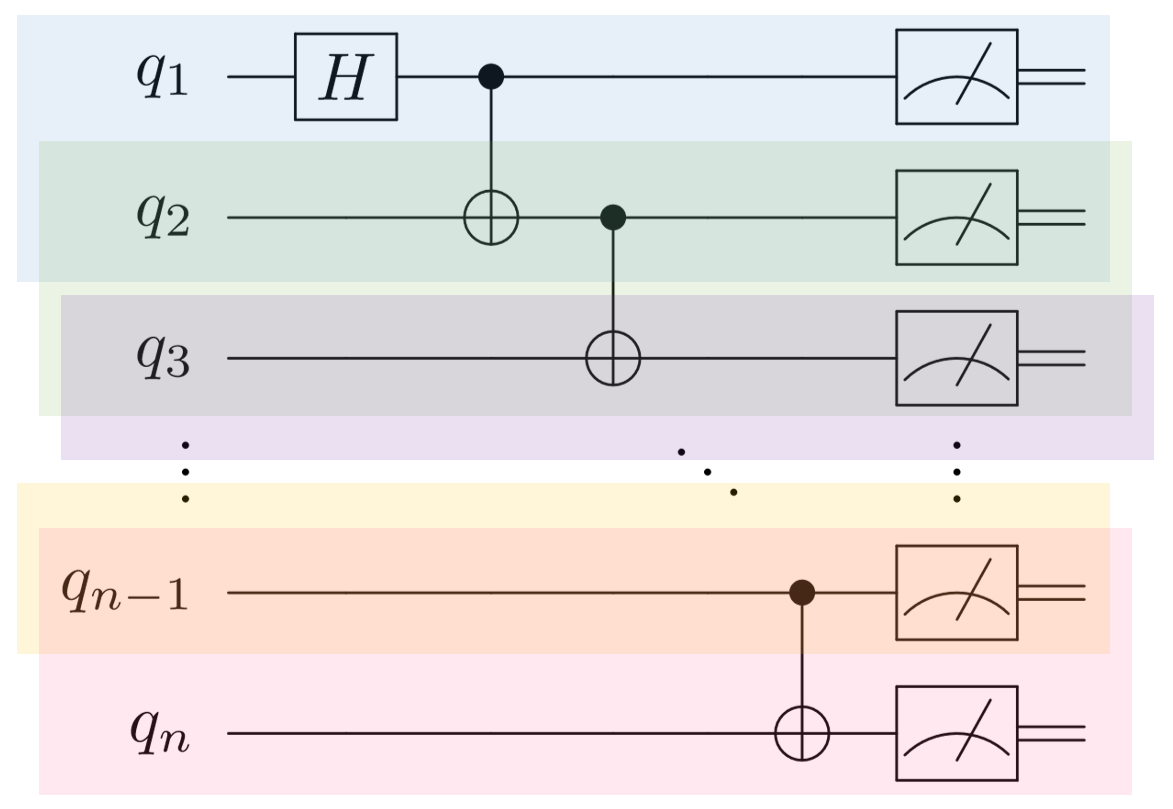}
\caption{The schematic representation of the quantum circuit used for preparation of the $n$-qubit GHZ state defined by \eqn{eq:ghzstate}. The circuit layout satisfies the connectivity constraints of the IBM \poughkeepsie QPU shown in Fig.~\ref{fig:layout}. The circuit uses a total of $n-1$ \cnot gates and $n$ measurement gates. Colored boxes denote subcircuit selections.}
\label{fig:GHZstatediagram}
\end{figure}

We consider the example of preparing the $n$-qubit GHZ state
\begin{equation}
    \ket{\textrm{GHZ}(n)} = \frac{1}{\sqrt{2}}\left(\ket{0_1,0_2,...,0_n} + \ket{1_1,1_2,...,1_n}\right)
    \label{eq:ghzstate}
\end{equation}
where the subscript denotes the qubit and the schematic representation of the input circuit $C$ is given in Fig.~\ref{fig:GHZstatediagram}. The instruction set for this circuit is limited to the one-qubit Hadamard $(H)$ and two-qubit controlled-NOT (\textsc{cnot}) unitaries along with the initialization and readout gates acting on a quantum register of size $n$. We study this example for a range of register sizes from $n=2$ to $20$ by composing a noisy circuit model that represents GHZ-state preparation on a QPU based on superconducting transmon technology \cite{chow2011simple,PhysRevA.93.060302}. This example demonstrates the unique features of superposition and entanglement using a circuit depth that is within the capabilities of the NISQ devices \cite{cruz2019,wei2019verifying}. 
\par
We decompose the GHZ-state preparation circuit from Fig.~\ref{fig:GHZstatediagram} into a set of subcircuits $S$ based on the procedure detailed in Sec.~\ref{sec:methods}. In this example, we identify a series of overlapping 2-qubit subcircuits for coarse-graining the $n$-qubit state preparation. Spatial variability in the device noise motivates a decomposition based on each register element $q_i$. We extend these subcircuits to generate a corresponding set of test circuits $T$ by the set $g$ given as 
\begin{equation}
    g=\{H_{q_0}, \cnot_{q_0,q_1}\}
\end{equation}
from which we select
\begin{equation}
    g'=\{H_{q_0}, H_{q_0} \circ \cnot_{q_0,q_1}\}
\end{equation}
The expected outcomes of these particular test circuits are simple to calculate from the truth tables for each operator \cite{nielsen_chuang_2010}. We examine the models using these test circuits.
\subsection{Noisy Measurement Model}
We begin by characterizing the initialization and measurement test circuits, which are necessary for modeling noisy unitary gate behavior. The measurement process for each register element discriminates an analog signal to generate a classical bit \cite{PhysRevLett.114.200501}, and errors in signal discrimination may lead to the wrong value. Characterization of measurement records the number and type of outcomes observed for each initial state. We characterize each register element with respect to both the $0$ and $1$ output states.
The leading errors in the observed results occurs when the $j$-th register element maps an expected output value to its complement, i.e., $0\rightarrow 1$ and $1\rightarrow 0$.
\par 
We model measurement of the $j$-th element as a binary process 
subject to errors which act on the post-measurement classical bit string, and we consider two models for the measurement error process: symmetric readout noise (SRO) and asymmetric readout noise (ARO). The SRO model is defined by a single parameter $p_{\textrm{sro}}$ that specifies the probability for a bit to flip, and we define a test circuit to characterize this process as measurement immediately after initialization to state $\ket{0}$. We directly estimate the value of $p_{\textrm{sro}}$ from the number of errors when preparing this computational basis state as $p_{\textrm{sro}}=r(1)$, where $r(k)$ is the observed probability of $k$ errors recorded. This model implicitly delegates initialization errors to the readout error model. The SRO model is developed by test circuits $T=\{T_{meas}(\ket{0})\}$ where the final SRO model is defined by $M_{SRO} = M(T_{meas},p_{sro})$.
\par
By contrast, the ARO model uses two parameters: $p_0$ for the probability of error in readout of $\ket{0}$ and $p_1$ as the probability of error in readout of $\ket{1}$. The ARO model therefore represents a refinement of both the noise model parameters $p_i$ and the test circuit suite $T$. We may estimate $p_0$ using the same test circuit above, but we must extend the characterization to preparation and measurement of $\ket{1}$ to estimate $p_{1}$. These additional test circuits will require inclusion of the single-qubit $X$ gate, and we also add a test circuit for the $XX$ operation of two successive $X$ gates applied to a single qubit. The latter reproduces the initial state $\ket{0}$, enabling the error in readout of state $\ket{1}$ to be isolated from the error associated with the $X$ gate. The ARO model is therefore defined by $M_{ARO}=M(T,p_0,p_1)$ where $T=\{T_{meas}(\ket{0}),T_{meas}(\ket{1}),T_{XX}(\ket{0})\}$.
\par
We model the test circuits for the ARO process using  an isotropic depolarizing channel parameterized by $p_{\textrm{x}}$ to describe noise in the $X$ gate,
\begin{equation}
    \epsilon_{DP}(\rho) = (1-p_{x})I\rho I + \frac{p_{x}}{3}(X\rho X + Y\rho Y + Z\rho Z)
\end{equation}
where $I$, $X$, $Y$, and $Z$ are the Pauli operators. Characterization of the ARO model yields an overdetermined system of equations relating the four experimentally observed probabilities $r^{(X)}(0)$, $r^{(X)}(1)$, $r^{(XX)}(0)$, and $r^{(XX)}(1)$ to the parameters $p_0$, $p_1$, and $p_{\textrm{x}}$. Of these parameters, only the latter two are unknown since $p_0$ is determined by the same method outlined above for $p_{SRO}$. Because the experimental observations directly relate to each other via $r^{(X)}(0) + r^{(X)}(1) = 1$ and $r^{(XX)}(0) + r^{(XX)}(1) = 1$, we select the following system of equations for each register element based on counts of $r^{(\cdot)}(0)$.
\begin{equation}
    r^{(X)}(0) = \frac{2p_{\textrm{x}}}{3}\Big(1-p_0\Big) + p_1\Big(1-\frac{2p_{\textrm{x}}}{3}\Big)
\label{eq:readoutsyseqs1}
\end{equation}
\begin{equation}
\begin{split}
    r^{(XX)}(0) = & (1-p_0)\left[\left(1-\frac{2p_{\textrm{x}}}{3}\right)^2 + \left(\frac{2p_{\textrm{x}}}{3}\right)^2 \right] \\ 
    & +  p_1\left[\frac{4p_{\textrm{x}}}{3}\left(1-\frac{2p_{\textrm{x}}}{3}\right)\right]
\end{split}
\label{eq:readoutsyseqs2}
\end{equation}
This system of equations is solved using the SciPy function \texttt{fsolve}, which is based on Powell's hybrid method for minimization \cite{virtanen2019scipy}.
\subsection{Noisy Subcircuit Models}
Test circuits for characterizing noisy subcircuits generate results that include measurement noise. We use the noisy measurement model above to account for these behaviors when modeling the results from test circuits. For the SRO and ARO models discussed above, this directly estimates the probabilities expected to be observed for each register. We use this procedure when discussing the characterization below.
\par 
We first characterize the subcircuit representing the Hadamard operation. The test circuit for a single Hadamard is defined with respect to the expected values for input states drawn from the computational basis, which yield a uniform superposition of binary results upon ideal measurement. We also use even-parity sequences of Hadamard gates as a second test to estimate noise in the subcircuit. These test circuits $T=\{T_H(\ket{0}),T_{HH}(\ket{0}),T_{4H}(\ket{0}),T_{6H}(\ket{0}),...,T_{nH}\}$ are used to characterize the Hadamard gate to yield $M_H(T,p_H)$. 
\par
\begin{figure}[ht]
    \centering
    \includegraphics[width=0.25\textwidth]{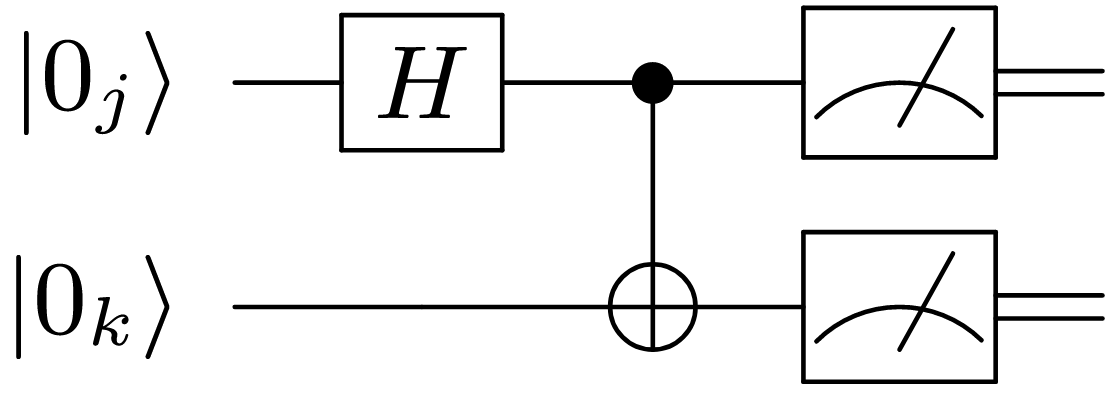}
\caption{The test circuit for characterizing the \cnot operation acting on register elements $q_j$ and $q_k$. This test prepares the two-qubit Bell state as an instance of $n=2$ in Fig.~\ref{fig:GHZstatediagram}.
}
\label{fig:CNOTnoise}
\end{figure}

We define test circuits for the \cnot operations that mirror the subcircuits used in the target application. For GHZ-state preparation, these are based on characterization of Bell-state preparation. The test circuit specification shown in Fig.~\ref{fig:CNOTnoise} produces the idealized result of a uniform distribution over perfectly correlated binary values. These test circuits may be defined across all pairings of register elements as represented by Fig.~\ref{fig:GHZstatediagram}. In particular, additional \cnot test circuits may be added to the set $g'$ from the set $g$, and additional \cnot test circuits for couplings not explicitly in $\mathcal{G}$ may be added as well. For convenience, we will denote the Bell-state preparation subcircuit as $U^{\textrm{BS}}_{(j,k)} = U_{(j,k)}^{(\cnot)} H_{(j)}\ket{0_j,0_k}$.
\par 
The noisy test circuits for Bell-state preparation are modeled by a pair of identical, independent depolarizing channels. Each channel, together defined as $\epsilon^{\textrm{DP}}_{j,k} = \epsilon^{\textrm{DP}}_{j} \otimes \epsilon^{\textrm{DP}}_{k}$, is parameterized by $p_{\cnot}$, which represents the probability of a depolarizing error determined independently for each qubit in the two-qubit \cnot gate. We therefore use the test circuit $T=\{T^{\textrm{BS}}_{(j,k)}(\ket{0_j,0_k})\}$ to compose model $M_{\cnot}=M(T,p_{\cnot})$. 
\par
The probability of observing bits $a$ and $b$ is given by 
\begin{equation}
\label{eq:belldensitymatrix}
    r_{j,k}(ab)=\textrm{Tr}\Big[\Pi_{ab}\epsilon^{\textrm{DP}}_{j,k}\left(U^{\textrm{BS}}_{(j,k)}\ket{0_j,0_k}\bra{0_j,0_k}U^{\textrm{BS}\dag}_{(j,k)}\right)\Big]
\end{equation}
where the operator $\Pi_{ab}$ projects onto the state $\ket{a,b}$, and the resulting trace yields the probability of the ideal measurement. 
The probabilities expected from the noisy Bell state subcircuit on qubits $j,k$ with ideal measurement is then given by 
\begin{align}
\label{eq:Bellidealmeas}
    r_{j,k}(00) = r_{j,k}(11) &= \frac{1}{2} - \frac{2}{3}p_{\cnot} + \frac{4}{9}p_{\cnot}^2 \\ \nonumber
    r_{j,k}(01) = r_{j,k}(10) &= \frac{2}{3}p_{\cnot} - \frac{4}{9}p_{\cnot}^2
\end{align}
Errors in readout transform these probabilities according to the noisy process, which may be either the SRO or ARO model.  
For example, the probability following readout $s_{j,k}(00)$ under the ARO channel is given by 
\begin{align}
\label{eq:cnotwithRO}
s_{j,k}(00) =& (1-p_0^j)(1-p_0^k) r_{j,k}(00) \\ \nonumber
&+ (1-p_0^j) p_1^k r_{j,k}(01) \\ \nonumber
&+ p_1^j (1-p_0^k) r_{j,k}(10) \\ \nonumber
&+ p_1^j p_1^k r_{j,k}(11)
\end{align}
\par
From the system of four equations generated by the readout probabilities $s_{j,k}(cd)$, we use the method of least squares to estimate $p_{\cnot}$. We minimize the sum of the squared residuals,
\begin{equation}
\sum_{cd} \Big(s_{j,k}(cd) - h_{j,k}(cd)\Big)^2
\label{eq:leastsquares}
\end{equation}
where each residual is defined as the difference between the modeled probability $s_{j,k}(cd)$ and the experimentally observed probability $h_{j,k}(cd)$ for each state result $cd$. The value $h_{j,k}(cd)$ represents the counts of state $cd$ on qubits $j,k$ measured during a total number of experiments $N_s$. The value returned for $p_{\cnot}$ is found using the SciPy \texttt{fsolve} function and bounded between 0 and 1 \cite{virtanen2019scipy}. 
\section{Experimental Characterization}
\label{sec:qpus}
\par 
In this section, we report on the results of experimental characterization and noisy circuit modeling of GHZ-state preparation using a QPU based on superconducting transmon technology developed by IBM. The IBM \poughkeepsie device has a register of 20 superconducting transmon elements that encode quantum information as a superposition of charge states \cite{koch2007charge}. 
Microwave pulses drive transitions between the  possible charge configurations and induce single-qubit gates. Coupling between register elements uses a cross-resonance gate that drives a mutual transition between transmons and therefore only occurs between two spatially connected elements \cite{chow2011simple}. 
\par 
The layout of the 20-qubit register in \poughkeepsie at the time of data collection is shown in Fig.~\ref{fig:layout}. A common edge in the connectivity diagram specifies those register elements that may interact through the cross-resonance operation. Individual registers are measured through coupling to a readout resonator, which results in a state-dependent change in the resonator frequency. Amplification of the readout signal then enables discrimination of the state using a quantum non-demolition measurement  \cite{corcoles2015demonstration, Gambetta2017}.
\par
Circuits are sent to the backend where they are translated into the appropriate ISA. The ISA for \poughkeepsie consists of the gates $U_1$, $U_2$, $U_3$, $CX$, and $ID$ \cite{openQASM}. The $U_1$, $U_2$, and $U_3$ gates are unitary rotation operators, of which $U_1$ is a ``virtual'' gate performed in software and $U_2$ and $U_3$ are performed in hardware. The identity gate $ID$ is used as a placeholder to create a timestep since it does not alter a quantum state. $CX$ represents the \cnot gate \cite{Qiskit-Textbook}. These instructions are implemented using low-level hardware operations. For instance, the $CX$ operator is implemented in hardware using a sequence consisting of cross-resonance gates and single-qubit rotation gates \cite{openQASM,2017CR,PhysRevB.81.134507}.
\par 
The \poughkeepsie QPU is accessed remotely using a client-server interface. We employ the Qiskit programming language to specify the input circuit and test circuits for the GHZ-state preparation application \cite{qiskitdocs}. These Pythonic programs are transpiled to the specifications and constraints of the backend, including ISA, connectivity layout, and register size. Additional inputs to the transpiler may include optimization protocols for minimizing circuit operations or noise levels. The transpiled programs are executed remotely on the \poughkeepsie device, which returns the corresponding measurements along with job metadata. 
\par
We use a shot count of 8,192 for all of the circuits executed on \poughkeepsie which represents the number of times each circuit is individually executed and generates the distribution of output states from the input circuit. Therefore each probability estimated by experiment is given by $r(k) = C(k)/N_s$, where $C(k)$ is the number of events observed for each measurement and $N_s$ is the shot count of 8,192. These measurements are subject to error due to variability in sampling in experiment from the QPU distribution. We restrict our sample size to a single experiment of 8,192 shots to avoid introducing effects from drift in the \poughkeepsie QPU. We use the standard deviation of these measurements to report error and statistical fluctuations, which is given by $\sqrt{(p(1-p)/N_s)}$ where $p$ is the binomial distribution probability parameter measured from experiment.
\par
\begin{figure}[t]
    \centering
    \includegraphics[width=0.45\textwidth]{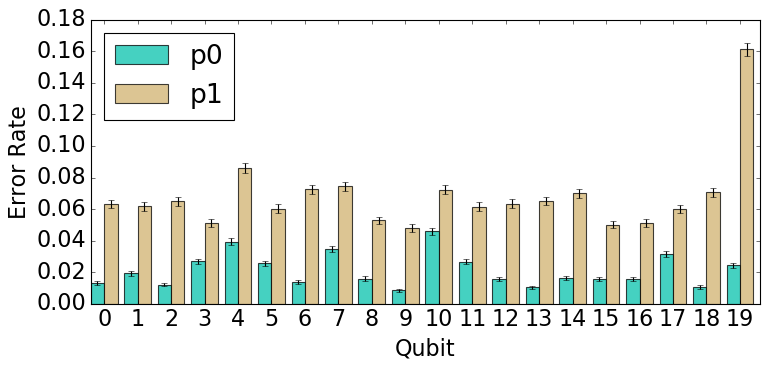}
    \caption{Error rates under the ARO channel for each qubit of \textit{poughkeepsie}. The SRO channel is given by the error rates for state 0 shown here. Average $p_0$ value is 0.0212 (standard deviation of 0.0101 across all qubits) and average $p_1$ value is 0.0681 (standard deviation of 0.0233). Each qubit is evaluated in a separate circuit, e.g. $X_0\ket{0_0,0_1,...,0_{19}}$ to generate Eq.~\ref{eq:readoutsyseqs1} for qubit 0.}
    \label{fig:readoutrates}
\end{figure}

We characterize measurement of all register elements in \poughkeepsie and analyze the results using the SRO and ARO models. The results for direct estimation of the ARO model parameter $p_0$ and $p_1$ are shown in Fig.~\ref{fig:readoutrates}. The results for the SRO model correspond with $p_{\textrm{sro}} = p_0$. From these results, we observe a large spatial variability in readout error as well as asymmetry per register element. The readout of state $\ket{1}$ is almost always more error-prone than readout of state $\ket{0}$.
\par
\begin{figure}[ht]
    \centering
    \includegraphics[width=0.45\textwidth]{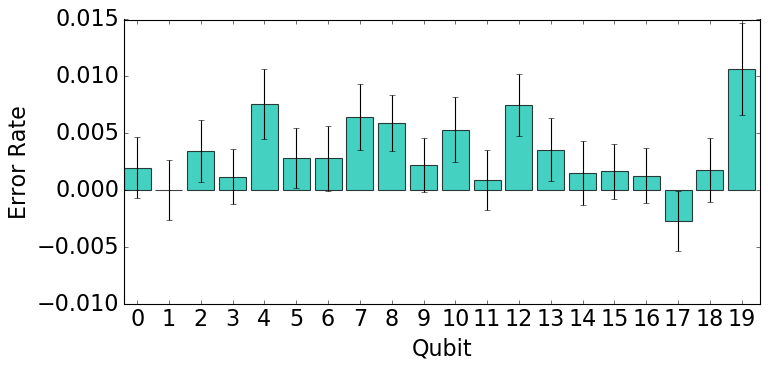}
    \caption{Depolarizing error rates associated with $X$ gate application for each qubit of \textit{poughkeepsie}. Average $p_x$ value is 0.0033 with standard deviation 0.00303.}
    \label{fig:xerrrates}
\end{figure}

 
The results of estimating the parameter $p_{x}$ for the depolarizing noise model of each $X$ gate are  shown in Fig.~\ref{fig:xerrrates}. From these results, we see spatial variability in the recovered error parameter. We observe one case of a negative error rate for qubit $17$ recovered from direct estimation using Eqs.~\ref{eq:readoutsyseqs1} and \ref{eq:readoutsyseqs2}. Because an estimated error rate of zero is within the experimental error, this is most likely due to statistical fluctuations. However, it could also be attributable to inconsistencies in the error behavior for the test circuits such that the model cannot estimate a feasible parameter based on the results, or to errors for this register that are not well described by a depolarizing channel such that a different model may yield a better solution. All other error rates are relatively small and therefore we have not investigated model refinement for this case because of the negligible contribution to the noise.
\par
We next characterize the Hadamard gate. We characterize error rates using test circuits generated from long sequences of Hadamards acting on a single element. We observe small error rates which correspond on average to 0.1\% error per gate.  
We attempted to model the Hadamard noise using a depolarizing channel but it did not lead to a better TVD than using a noiseless model for the gate. 
\par
We also characterized gate error models based on unitary rotation noise in $X$, $Y$, and $Z$ for the Hadamard gate which represents coherent errors. These characterizations did not yield a smaller TVD than using a noiseless model. Our choice to restrict characterizations to computational basis measurements significantly limits the achievable accuracy or effectiveness of this model. In general, such characterizations are not capable of identifying arbitrary coherent noise and are limited, e.g. only $X$ and $Y$ noise have an observable effect in the $Z$ measurement basis. Additional test circuits could address this limitation at the expense of increased experiment count. For our purposes, we concluded that error rates associated with the Hadamard operation were negligible as this noise was 100 times smaller than the next leading gate error.
\par
\begin{figure}[ht]
    \centering
    \includegraphics[width=0.45\textwidth]{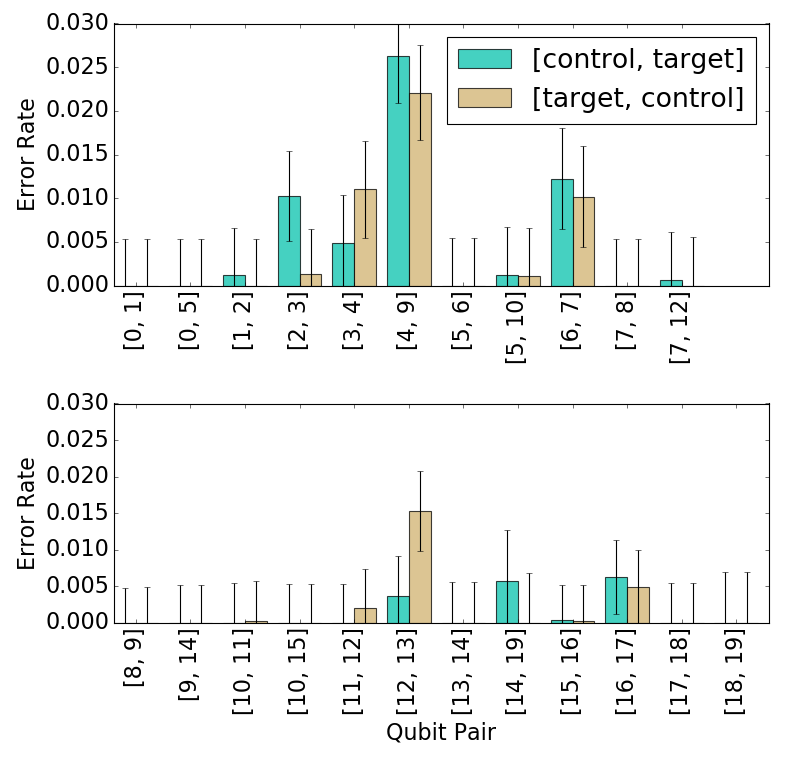}
    \caption{Error rates for \cnot gates under the depolarizing channel for each coupled qubit pair of \textit{poughkeepsie}. These values are fitted to include the ARO channel noise with rates shown in Fig.~\ref{fig:readoutrates}. Reported error bars represent the upper limit of the error from the least squares calculation.}
    \label{fig:cnot1}
\end{figure}

 
We next characterize the Bell-state preparation circuits for each pair of possible interactions shown in Fig.~\ref{fig:layout}. We select the depolarizing noise model because it is a well-understood model for quantum noise that captures several different fundamental aspects of quantum behavior. We do not expect the depolarizing model to be a perfect fit to experimental data but this model provides a useful method to understand noise levels in the system and how noise from different components interacts. We use least-squares error estimation to find the value of depolarizing parameter $p_{\cnot}$ that best fits the results while accounting for readout error as in \eqn{eq:cnotwithRO}. This approach yields more consistent results than solving each equation in the system explicitly and using a selection process to determine the final $p_{\cnot}$ value from among these solutions which are often highly varied. The estimated parameter values are shown in Fig.~\ref{fig:cnot1}. The magnitude of the error bars for the parameter estimations highlights the relative magnitude of gate noise to readout noise.
\par
We test the accuracy of the noisy subcircuit models with estimated parameters from experimental characterization. For these tests, we use explicit numerical simulation of the quantum state prepared by each noisy subcircuit model. We estimate the measurement outcomes for these modeled circuits using the simulated quantum state, and we compare these simulated observables with the corresponding experimental observations from the \poughkeepsie device. The accuracy of the noisy subcircuit model is quantified using the total variation distance (TVD) defined in \eqn{eq:TVD}.
\par
Our simulations of the quantum state use a numerical simulator bundled into the Qiskit software framework. The Aer software simulates both noiseless and noisy quantum circuits using the same Qiskit programs sent to the \poughkeepsie device as input. We constrain the simulator to a statevector simulation method. Within Aer, we input the noise models using the error rates and noise operators of depolarizing and readout channels as defined in Sec.~\ref{sec:application}. Aer models gate noise using error functions parameterized by these error rates which create noisy descriptions of gates for simulation. When a noisy simulation is run, these functions sample errors and inject them as operations within the circuit. We tailor the simulations to match the developed noisy subcircuit models. Each test case acquired $N_s$ samples in order to mimic the finite statistics from experimental characterization. We generate a number of simulation samples of 8,192 shots per sample to create a sampling distribution. We report the standard deviation of this distribution which represents error due to variability in sampling in simulation.
\par
\begin{figure}[b]
    \centering
    \includegraphics[width=0.48\textwidth]{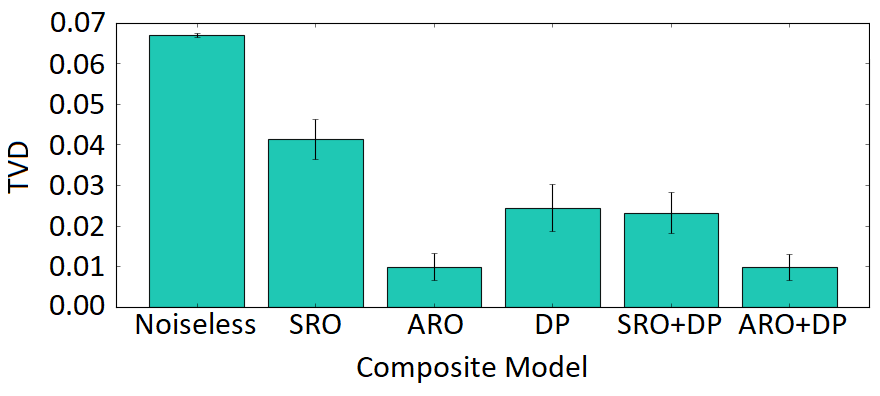}
    \caption{Comparison of possible choices for composite model. The best performance is achieved in the ARO+DP case. Error bars represent the distribution of TVD values across 100 sets of 8,192 samples per simulation case.}
    \label{fig:compositemodels}
\end{figure}

A comparison of accuracy for different noisy subcircuit models is shown in Fig.~\ref{fig:compositemodels} for simulating the Bell state circuit on qubits 0 and 1 on the \poughkeepsie device. We calculate the TVD between experiment and simulation using six different noise cases. We consider symmetric readout only (SRO), asymmetric readout only (ARO), \cnot depolarizing error only (DP), symmetric readout with \cnot error (SRO+DP), and asymmetric readout with \cnot error (ARO+DP). The error rate parameters are optimized for each composite noise model, e.g.~the optimal depolarizing parameter in the SRO+DP case may not be the same value found for the ARO+DP case. We also simulate a noiseless Bell state for a baseline comparison. 
\par
The results shown in Fig.~\ref{fig:compositemodels} clarify the noisy circuit model yielding the smallest TVD is composed from the asymmetric readout channel with a \cnot depolarizing channel (ARO+DP). Since each noise model achieves a clear improvement in TVD as measured by a decrease from the noiseless case that is outside of error bars, we can be confident that each selected model is capturing some of the noise behavior present in the system while also illustrating which models provide the best descriptions of the noise. For example, in the noise model case `DP' we have modeled a depolarizing channel for which the $p_{\cnot}$ parameter is calculated to account for all noise in the system. This model has a clear improvement on TVD and therefore is likely to be an effective description of the noise in the system. However, the addition of readout noise models for the `SRO+DP' and `ARO+DP' cases is evidently a more accurate noise model because these models achieve further improvements in TVD.
\section{Performance Testing Results}
\label{sec:performance}
\begin{figure}[htbp]
    \centering
    \includegraphics[width=0.45\textwidth]{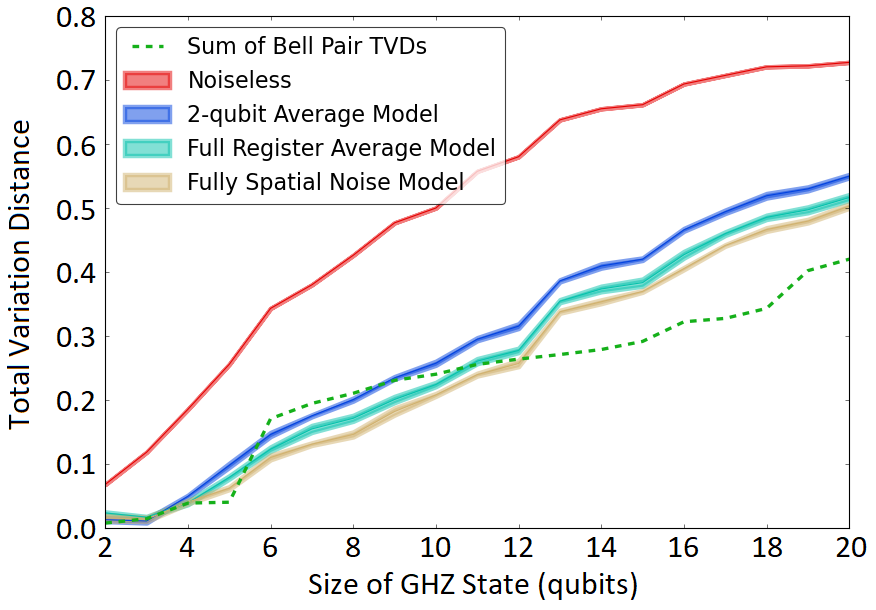}
    \caption{Performance of selected noise model on $n$-qubit GHZ states. The best performance is achieved with the fully spatial noise model. Error bars represent the distribution of TVD values across 6 sets of 8,192 samples per simulation case.}
    \label{fig:ghzperformance}
\end{figure}

We now present the performance of the selected composite model on $n$-qubit GHZ-state preparation circuits. Using the estimated ARO and \cnot error rates, we demonstrate iterations of this composite noise model which represent varying model complexity and experimental efficiency to achieve a particular accuracy. These iterations are shown in Fig.~\ref{fig:ghzperformance}. The 2-qubit average case represents the performance of a noise model with only three parameters--$p_0, p_1, p_{\cnot}$--which are taken as the average of the error rates for only qubits 0 and 1. This represents a case of characterization using the fewest quantum resources, requiring only 7 experiments. We also consider a case which uses these same three parameters averaged over the entire register which retains low model complexity of only three noise parameters but requires the full suite of experiments. Our most detailed model accounts for spatial variations in the error parameters and uses individualized readout error rates for each qubit and  \cnot error rates for each coupling. As with the Bell state example in Fig.~\ref{fig:compositemodels}, we show the noiseless case for the sake of context and comparison. Finally, we also show the sum of the minimum TVD achieved for noisy simulation of the Bell state across each qubit pair for which a \cnot was applied in the GHZ preparation circuit.
\par
\begin{figure}[htbp]
    \centering
    \includegraphics[width=0.45\textwidth]{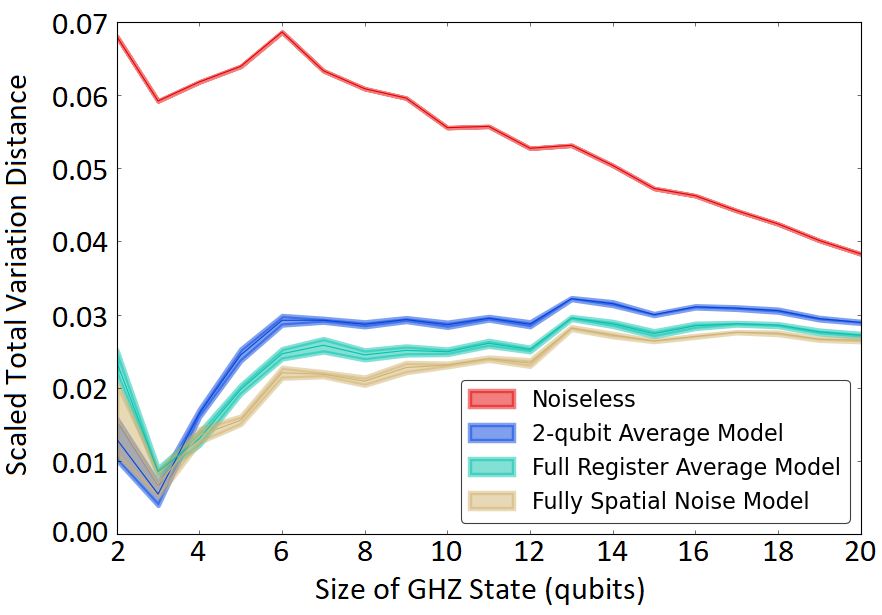}
    \caption{Scaled performance of selected noise model on $n$-qubit GHZ states, where TVD is divided by the number of \cnot gates in each circuit. Error bars represent the distribution of TVD values across 6 sets of 8,192 samples per simulation case.}
    \label{fig:ghzscaled}
\end{figure}

Figure \ref{fig:ghzperformance} demonstrates a significant improvement in model accuracy for GHZ state preparation using our composite noisy circuit model. The improvement is a 3-fold decrease in TVD as compared to the noiseless simulation. Our fully spatial model performs better than the coarser-grained models, such as the average two-qubit model, particularly for larger sizes of GHZ state preparation. We also examine the scaling in the error with respect to the area of the circuit. We normalize the computed TVD by the number of \cnot gates in each GHZ preparation circuit, and we find that the per-qubit model accuracy is nearly constant across all GHZ circuit instances, as shown in Fig.~\ref{fig:ghzscaled}. This trend would also hold when TVD is scaled by qubit count, since qubit count and \cnot count are strongly linked in the GHZ example. Since the TVD increases at a rate commensurate with \cnot count or qubit count, this may indicate that higher levels of entanglement or larger Hilbert spaces impact the predictability of noise in the device. 
\section{Bernstein-Vazirani Application}
\label{sec:bv}
\begin{figure}[ht]
    \centering
    \includegraphics[width=0.45\textwidth]{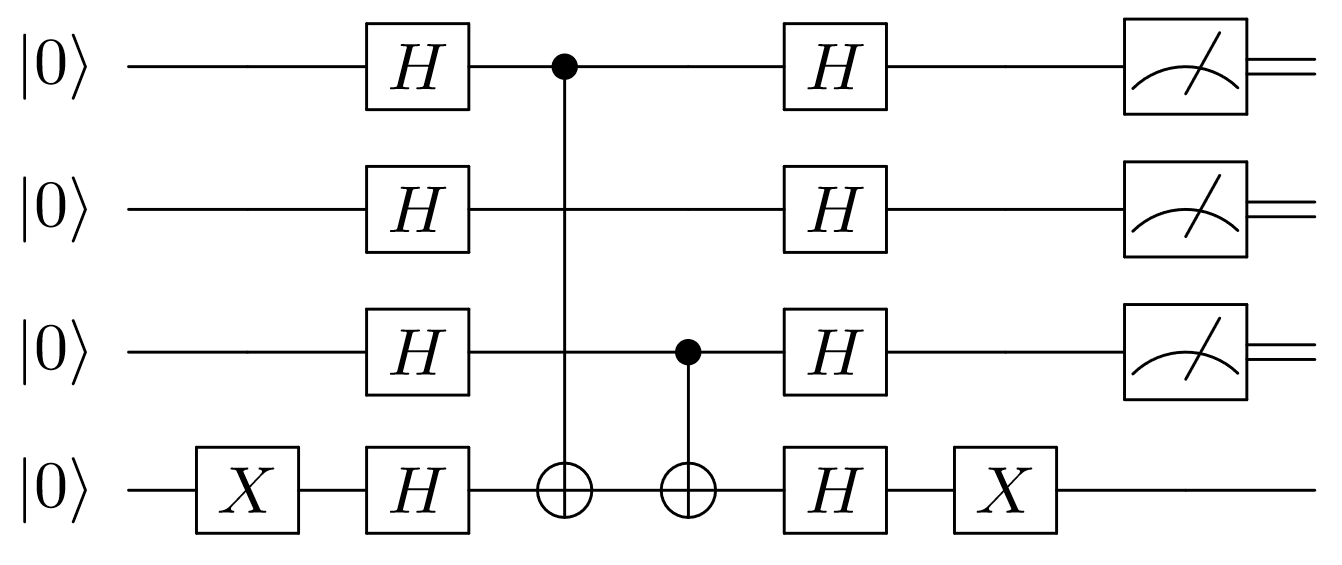}
    \caption{Circuit implementation of the Bernstein-Vazirani algorithm. The bottom qubit of the register is the oracle; the top three yield the secret string, here given as 101 as example. Other secret strings are produced by changing the \cnot gate sequence such that control qubits correspond to output bits of 1.}
    \label{fig:bvcircuit}
\end{figure}

We next test the performance of this noisy circuit model on a different application to evaluate its ability to capture fundamental characteristics of the device. We test the performance by modeling several quantum circuit instances of the Bernstein-Vazirani algorithm. This algorithm considers a black box function that is encoded by a secret binary string which the Bernstein-Vazirani algorithm finds in one query \cite{BValgorithm}. Figure \ref{fig:bvcircuit} shows an example of our circuit implementation of this algorithm using a three-bit string. We use a phase oracle qubit as the black box function encoded with the secret string. Upon measurement of the non-oracle qubits we obtain the secret binary string. We select the Bernstein-Vazirani algorithm because it is implemented using the same gate set we have characterized for the GHZ example, so we do not require additional characterization circuits. 
\par
Given the connectivity constraints of the \poughkeepsie device, the maximum bit string we can test without introducing SWAP operations is of length three. We choose qubits 6, 8, and 12 with oracle qubit 7 because this set has among the lowest error parameters. We execute the Bernstein-Vazirani algorithm for every possible encoding of the three-bit secret string and record the accuracy as the probability that the encoded string was observed. We include collection of these measurements during the same job used to characterize the device.
\par
\begin{figure}[t]
    \centering
    \includegraphics[width=0.45\textwidth]{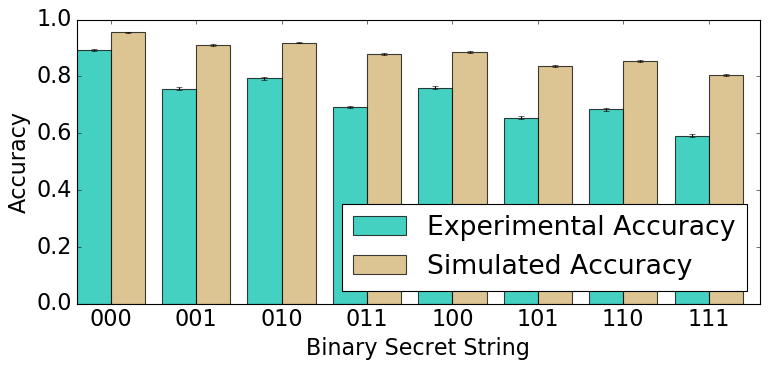}
    \caption{Performance of Bernstein-Vazirani algorithm evaluated as the measured probability of the prepared secret string. Simulation is subject to noise defined by the fully spatial model.}
    \label{fig:bvdual}
\end{figure}

 
Figure \ref{fig:bvdual} plots the simulated accuracy of the circuit outcome using the fully spatial noise model alongside the experimental accuracy. Our model captures the decrease in experimental observed accuracy across the various binary strings. The loss in accuracy scales with the number of 1 bits in the secret string for both the experiment and simulation. However, the accuracy predicted by simulation is consistently higher than the accuracy observed experimentally, indicating a state-dependent noise source remains missing from this model. 
\section{Conclusion}
\label{sec:fin}
We have presented an approach to noisy quantum circuit modeling based on experimental characterization. Our approach relies on composing subcircuit models to satisfy a desired accuracy threshold, model complexity, and experimental efficiency, which we implement using the total variation distance. We have tested our ideas using the IBM \poughkeepsie device, which enables evaluation of our characterization methods as well as the comparison of predicted performance for GHZ-state preparation and an instance of the Bernstein-Vazirani algorithm. The initial example focused on GHZ-state preparation examined model fidelity with respect to both width and depth of an input circuit. Models for the readout and \cnot subcircuits accounted for a majority of the  model error. Our analysis of a second test circuit using instances of the Bernstein-Vazirani algorithm reveals additional sources of errors not captured in the original GHZ circuit characterization. Because both tests depend on the same gates for state preparation, the appearance of new errors suggests a possible state-dependent noise model that warrants further investigation. While our demonstrations have focused on specific devices and input circuits, the methodology provides a robust and flexible framework by which to generate noisy quantum circuit models on any device.
\par
A significant feature of this approach to noise model decomposition is to iteratively adjust the models until sufficient accuracy is obtained. Improvements in accuracy may be obtained by changing characterization circuits or parameter estimation. The Bell-state and GHZ-state preparation examples demonstrate how this model adjustment may be performed by varying the experimental efficiency and the input to the model to change the accuracy of the final composite model. Our demonstrations have focused on the depolarizing channel for gate modeling, but circuit characterization can be directly extended to account for new noise models, components, applications, and algorithms. For example, in both the GHZ and Bernstein-Vazirani results, we observe an increase in TVD that scales with the number of \cnot gates applied in the circuit. A more sophisticated \cnot noise model may improve accuracy of the final noise model. Since placing limitations on coarse-graining may introduce insensitivities to certain error types, for instance measurement only in the computational basis creates insensitivity to $Z$ error types, it will likely be necessary to refine test circuits to address more sophisticated models. Additionally, this methodology assumes separability in composition-decomposition, i.e.~it assumes that the noise present in the decomposed subcircuits is not substantially different from that of the composed circuit and that any differences may be tuned away by refinement. If this assumption is not true, there may be an upper limit to the achievable accuracy of noise modeling using subcircuit testing. Further model refinement and testing would be necessary to demonstrate this non-separability.
\par
Our original motivation was to address the growing challenge of characterizing NISQ applications, for which efficient and scalable methods are necessary. We have shown how to construct a set of test circuits that scales with the area of the input circuit $C$ and the underlying decomposition strategy. In the GHZ-state preparation example, the number of total experiments needed for full spatial characterization scales with the size of the register $q$ and the number of couplings $c$ according to $N_s(2q+2c+1)$. This resource requirement enables characterization to be run alongside the state preparation circuit when the job is sent to the QPU. This efficiency should help ensure noise characterization is performed within the same processor context as the sought-after circuit. We anticipate such real-time characterizations to be valuable for dynamic compiling and tuning of quantum programs \cite{Murali:2019:NCM:3297858.3304075,tannu2018case,10.1117/12.2526670}. 
\par
Our approach to characterization has relied on model selection using minimization of the total variation distance (TVD) between noisy simulation and experimental results. This demonstration used a small set of the possible models for characterizing the observed QPU behavior, and expanding the set of potential models is possible for future work. There is a necessary balance, however, between the sophistication of the model and the utility for characterizing QPU behavior. While fine-grain quantum physical models are capable of capturing a more detailed picture of the dynamics present on small scales, the dawning of the NISQ era requires the addition of new techniques to our toolbox that have a higher-level and larger-scale approach. For scalable numerical analysis of quantum computational methods, it is essential that we develop coarse-grained, top-down approaches to capture the core behavior of QPUs.
\section*{Acknowledgements}
{This research is supported by the Department of Energy Office of Science Early Career Research Program and used resources of the Oak Ridge Leadership Computing Facility, which is a DOE Office of Science User Facility supported under Contract DE-AC05-00OR22725.}

\bibliography{ref}
\bibliographystyle{plain} 
\end{document}